# Origin of crack pattern in the deposition from drying colloidal suspension

Jun Ma[1], Guangyin Jing[1,2]
[1]Department of Physics, Northwest University
[2]NanoBiophotonics center, National Key Laboratory and Incubation Base of Photoelectric Technology and Functional Materials, Xian, 710069, China

The fracture mechanics was widely employed to explain the crack propagation in the deposition produced by drying colloidal suspension. However, more complex than conventional fracture, those cracks periodically distribute and make up a unique pattern. This still remains mysterious so far. Inspired by the concept of spinodal decomposition, here, we develop the theory to elucidate the spatial arrangement of the cracks, which indicates that the crack pattern is generated by the phase separation of colloidal clusters and water. It concludes that the crack spacing results from the wavelength of the concentration fluctuation during the phase separation, linearly growing with the increase of the deposition thickness and initial particle concentration, which is consistent with experimental results.

Colloidal depositions formed by drying suspension ubiquitously occur in both nature and technology application. Cracks widely exist in those depositions, which greatly limits and lowers their practical purposes. More elusive than ordinary cracks in routine materials, cracks in colloidal deposition periodically distribute, and divide the 2-D deposit into uniform domains. These cracks separated by a distance make up the array pattern, which is characterized by the wavelength (crack spacing) [1-5].

Crack patterns displayed versatile morphologies under various experimental conditions. Drying suspension of silica or polystyrene spheres, Okubo *et al.* demonstrated the patterns consisting of spoke-like cracks [3, 4]. Polygon pattern of crack was produced after drying starch slurry [1, 5]. Shorlin *et al.* studied crack patterns formed by drying a thin layer of alumina/water slurry. The web-like crack pattern produced from isotropic drying, whereas the parallel longitudinal pattern formed on the condition of directional drying [2].

The fracture mechanics was employed to explain the crack propagation. Tirumkudulu and Russel studied cracks in dried latex films, and found the scaling law of critical stresses derived from the concept of classical Griffith's energy balance [6]. On the other hand, experimental results show that the periodical characters of crack pattern depend on the deposition thickness [7, 8] and particle concentration [4, 9], etc. The theory of those factors on the formation of crack pattern is still missing. Inspired by the work of spinodal decomposition [10], here, we propose a theory to explore the origin of the pattern consisting of periodically parallel cracks.

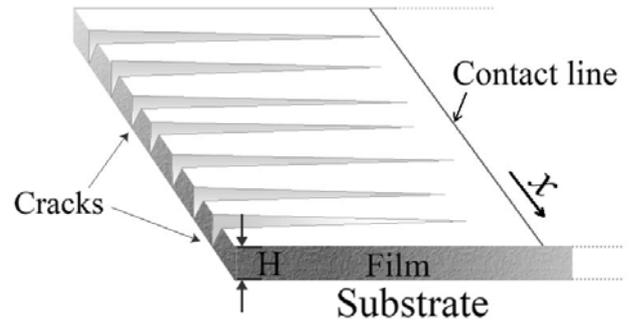

Fig. 1. Sketch of the crack pattern resulting from directional drying colloidal suspension on the rigid plate.

Consider the directional drying colloidal suspension (hard spheres + distilled water) on a rigid wettable plate (Fig. 1). Although the DLVO theory is valid for dilute suspensions, it is also applied for higher solid contents on the condition of only main characteristic such as interaction energy being concerned [11]. Since the dilute suspension concentrates gradually in the whole evaporation process, the DLVO theory is adopted to describe the interaction energy between two colloidal particles (clusters),

$$\phi = (\frac{2k_B T}{e}\ln\zeta + \psi_0) - \frac{AR}{12\zeta} \qquad (1)$$

$k_B$ is Boltzmann's constant, $T$ temperature (K), $e$ electron unit charge, $\zeta$ the surface distance of two particles (clusters), $\psi_0$ reference potential, $A$ Hamaker constant, $R$ particle (cluster) radius. The first term represents the electric potential derived from Poisson-Boltzmann equation on the condition of weak electrolyte (No added electrolyte) [12], and the second term is the van der waals energy [13].

The coalescence of colloidal particles results from the collision which is achieved by Brownian motion. To simplify the analysis, we assume the doublet-reaction to occur, which has been adopted for the general analysis of colloidal aggregation [14]. Due to weak electrolyte, the repulsive electric force is strong so that particles coalesce by reaction-limited colloid aggregation (RLCA).

During the evaporation time $t^e$, the contact line is not pinned and continually retreats (Fig. 2(a)), laying the deposit where it passes away, which indicates that no flow replenishes solvent loss owing to the evaporation. The particle diffusion from inner suspension to the reaction area can be neglected (see Appendix $A$). This is consistent with Roth $et$ $al.$ experimental study of drying colloidal droplet with retreating contact line, in which no particle transportation occurred by diffusion or flow from inner drop to the drying front (near contact line), and the particle deposition resulted from concentrating due to the drying [15].

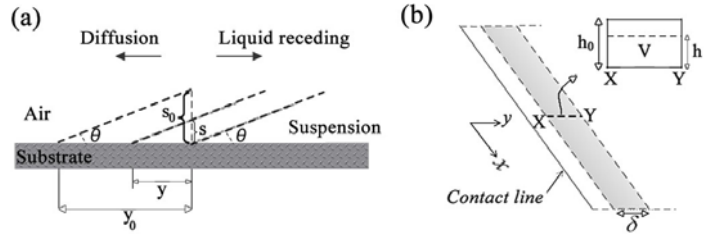

Fig. 2. (a) Sketch of suspension receding in drying process. (b) Strip area with the width δ close to the contact line. Inset, cross section of the strip area (section line $XY$).

We study the strip area near the contact line with the width δ (Fig. 2(b)). Since δ is short, the difference of suspension height along $y$ direction can be ignored (see the section figure along $XY$ in the inset). As indicated by the study of the interaction of colloid clusters, the cluster shape can be approximately treated as sphere [16]. We consider that the clusters continuingly grow and separate with a distance $\zeta$ in the solvent. This physical configuration is confirmed by the $X$-ray scattering experiment for drying nanoparticle suspension [15, 17]. The suspension volume can be represented as, $V = \int dV$, the variable of integration $dV = (2R+\zeta)^3$ (see Fig. 3(a), since $dV$ is very small, particles are approximately treated as locating on the lattice sites). For $N$ homogenously distributed particles during the $k$th coalescence process,

$$V = S(h_0 - h_t' t^e_{k-1}) = N(2R+\zeta)^3/4 \qquad (2)$$

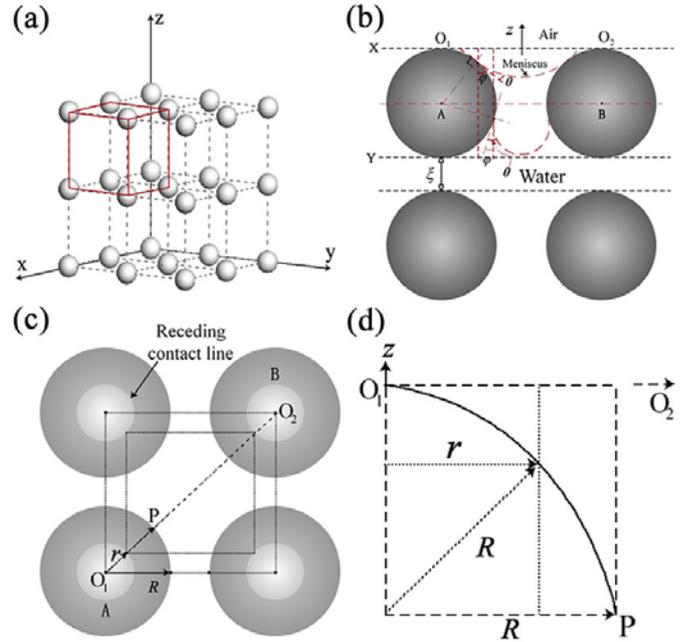

Fig. 3. (a) Cubic representation of clusters locating at lattice sites. (b) Side view of the drainage. (c) Top view of the drainage. Small bright circle is the projection of bare spherical cap produced by the drainage. (d) Section figure along $O_1P$ in (c).

where $S$ is the evaporation area, $t^e_{k-1}$ evaporation (reaction) time for completing $k-1$ times of coalescences. $h_0$ initial suspension height at $t^e = 0$, $h_t'$ ($\approx$ 0.1 μm/s [8]) evaporation rate along the height direction. After $k$ times of coalescences, the numbers and radius of clusters in the suspension are,
$N = N_0/2^k$ (3)
$R = 2^{k/3} r_0$ (4)
respectively; where $N_0$, $r_0$ are the numbers and radius of initial colloidal particles, respectively. Substitute equation (3) into (2), get the cluster distance,

$$2R+\zeta = \left[(\frac{2^{k+2}}{n_0})(1-\frac{h_t'}{h_0}t^e_{k-1})\right]^{\frac{1}{3}} \qquad (5)$$

where $n_0 = N_0/Sh_0 = N_0/V_0$, is the initial particle number concentration.

During the period of one coalescence ($t_k$), the concentration increment resulting from the evaporation is (Appendix *B*),

$$\Delta C_E = n_0 h'_t t_k / h_0 = n_0 t_k / t \qquad (6)$$

where $t$ is the total evaporation time. For one coalescence, the number concentration halves in term of equation (3), the corresponding concentration reduction $|\Delta C_C| \gg \Delta C_E$ (Appendix *B*). Moreover, equation (5) indicates that the cube of cluster distance grows in the power-law with the increasing of the times ($k$) of coalescences, whereas linearly decreases with the evaporation time $t^e$. Therefore, although the number concentration slightly increases with the evaporation, it greatly decreases due to the cluster aggregation. During one coalescence, we can ignore the evaporation effect and consider that the number concentration is controlled by cluster reaction. However, the accumulated effect of evaporation on the concentration in long time has achieved by shortening the distance of clusters, given by equation (5).

In the RLCA process, numbers of collisions lead to one coalescence. The coalescence probability of particle collision is, $P$ = numbers of coalescence / numbers of collision [18]. The characteristic time for doublet formation can be derived from the probability density theory [19]. Combining $P$ into this result, we get the time for completing one coalescence,

$$t_k = \frac{\pi \eta W R^3}{k_B T P v} \qquad (7)$$

where the stability ratio

$$W = \frac{2R}{r_m^2}\left(\frac{2\pi k_B T}{-(\partial^2 \phi / \partial r^2)_{r_m}}\right)^{1/2} e^{\phi_m / k_B T}, \phi_m \text{ the maximum}$$

potential evaluated by the derivative of equation (1), satisfying $(d\phi/dr)_{r_m} = 0$, $r_m$ (= $2R + \zeta_m$) the distance between the centers of two adjacent clusters (as $\phi = \phi_m$), $(\partial^2 \phi / \partial r^2)_m$ (negative value) the second derivative of $\phi$ at $r_m$. The cluster volume fraction

$$v = N \frac{4}{3}\pi R^3 / \frac{N}{4}(2R+\zeta)^3 = \frac{16\pi}{3}\left(\frac{R}{2R+\zeta}\right)^3, \text{ thus,}$$

$$t_k = \frac{3\eta W(2R+\zeta)^3}{16 k_B T P} \qquad (8)$$

Combining equations (5) and (8), the total time for finishing $k$ times of coalescences is,

$$t = \sum_{k=1}^{k} t_k = \frac{3\eta W}{16 k_B T P n_0}\sum_{k=1}^{k} 2^{k+2}(1-\frac{h'_t}{h_0}t^e_{k-1}).$$ As the $k$th coalescence proceeding, the evaporation time having been experienced since $t^e = 0$ is the sum of the time for $k-1$ times of coalescences, $t^e_{k-1} = \sum_{k=1}^{k-1} t_k$. With series expansion and approximation, we reach an equation, $t \approx \frac{2^k 3\eta W}{2 k_B T P n_0}(1-\frac{h'_t}{h_0}t)$. Solving the equation, we get the times (numbers) of coalescences before the phase separation of water and clusters,

$$k = \log_2 \frac{2 k_B T P n_0 h_0 t}{3\eta W (h_0 - h'_t t)} \qquad (9)$$

With water continuously evaporating, at the late stage of drying process, air invades the upper layer of clusters to form a meniscus (Fig. 3(b)). The capillary effect results in water drainage from the clusters, driving further cluster coalescence. Since the water flow is induced by the capillary force, thus the change of surface energy in unit volume is

$$E = (\gamma_c - \gamma_{wc}) n 4\pi R^2 = 4\pi \gamma n R^2 \qquad (10)$$

where $n$ is the numbers of clusters in unit volume, $\gamma_c$ and $\gamma$ are surface energies of cluster and water in air, respectively, $\gamma_{wc}$ the interfacial energy between water and cluster.

The maximum van der waals potential of the coalescence of clusters is no more than 70 $KT \sim$ 2.8 × 10$^{-19}$ J at room temperature [11], which is far less than the driving energy here for phase separation of $\sim$ 3.1 × 10$^{-11}$ J, released from the surface energy ($2\gamma 4\pi R^2/6$, as one spherical cluster can coalesce with 6 neighboring spheres) for a pair of typically coalescing clusters of 10 μm diameter. Therefore, the van der waals energy can be neglected in the process of phase separation.

The total work for water drainage in unit volume is (Appendix *C*),

$$W = \frac{4}{3}\pi n R (2R + \zeta)\cos\theta \qquad (11)$$

where $\theta$ is the contact angle of water on the spherical cluster. To meet the energy criterion of water drainage, $W - E \leqslant 0$, the surface distance of adjacent clusters satisfies,

$$\zeta \leqslant R(3\sec\theta - 2) \qquad (12)$$

During phase separating, water flow through the porous medium (clusters), described by Darcy's law as [20],

$$Q = \frac{-kA}{\eta}\frac{\partial(\Delta P)}{\partial x} \quad (13)$$

where $Q$ is the flow rate, $k$ permeability of the medium, $\partial(\Delta P)/\partial x$ pressure gradient (negative value), $A$ cross-sectional area perpendicular to $x$, and the pressure difference $\Delta P = \frac{\Delta G}{xA} = E - W$. Thus we have,

$$\Delta P = 4\pi\gamma nR(R - \frac{1}{3}(2R+\zeta)\cos\theta) \quad (14)$$

On the other hand, the flow rate can also be represented as,

$$Q = \frac{\partial V_w}{\partial t} = \frac{\partial(1-4\pi R^3 n/3)V_s}{\partial t} = -\frac{4\pi R^3 V_s}{3}\frac{\partial n}{\partial t} \quad (15)$$

where $V_w$ is the change of water volume due to the drainage of suspension volume $V_s$. Take the derivative of equations (13) and (15) with $x$, and combine with $\partial V_s/\partial x = A$ and equation (14), we get,

$$\frac{\partial n}{\partial t} = \frac{k\gamma((2R+\zeta)\cos\theta - 3R)}{\eta R^2}\frac{\partial^2 n}{\partial x^2} \quad (16)$$

This equation has the similar solution as that of Cahn's equation [10], which represents Fourier components of the fluctuation of cluster concentration due to inhomogeneous distribution,

$$n - n_c = \sum_{\alpha_i} Exp[q(\alpha_i)\tau]\cdot[B(\alpha_i)\cos\alpha_i x + K(\alpha_i)\sin\alpha_i x] \quad (17)$$

where $n_c$ is the numbers of homogenously distributed clusters in unit volume, $\tau$ time of the fluctuation, $\alpha_i$ wavenumber. $B$ and $K$ are evaluated at $\tau = 0$ by Fourier analysis. $q$ is amplification factor,

$$q(\alpha) = k\gamma\alpha^2(3R - (2R+\zeta)\cos\theta)/\eta R^2 \quad (18)$$

As the evaporation proceeding, $\zeta$ is continually decreasing until it satisfies $W - E \leq 0$, i.e. $3R - (2R+\zeta)\cos\theta \geq 0$. This leads to $q \geq 0$, i.e. the fluctuation amplifies with time. Thus the cluster suspension is unstable, leading to the separation of cluster and water similar to the spinodal decomposition [10]. As $q = 0$, the phase separation starts with the critical wave number α. Thus, equations (17) can be written as,

$$n - n_c = Exp[q(\alpha)\tau]\cdot[B(\alpha)\cos\alpha x + K(\alpha)\sin\alpha x] \quad (19)$$

Take the derivative of two sides of equation (2), we have,

$$(2R+\zeta)\frac{dN}{dx} + 3N\frac{d\zeta}{dx} = 0 \quad (20)$$

The coordination $x$ can be represented as, $x = N(2R + \zeta)$. Take its derivative $N\frac{d\zeta}{dx} = 1 - (2R+\zeta)\frac{dN}{dx}$, we rewrite equation (20) as,

$$(2R+\zeta)\frac{dN}{dx} = \frac{3}{2} \quad (21)$$

To evaluate $dN/dx$, multiply $V$ and take the derivative of equation (19) at $\tau = 0$, it gives,

$$\frac{dN}{dx} = \alpha V\sqrt{B^2 + K^2}(-\frac{B}{\sqrt{B^2+K^2}}\sin\alpha x + \frac{K}{\sqrt{B^2+K^2}}\cos\alpha x),$$

where $V\sqrt{B^2+K^2} = Vcn_c = cN_c$ ($N_c$, cluster numbers at $\tau = 0$), $c$ is decimal coefficient determined by the initial and boundary conditions. Make $\frac{K}{\sqrt{B^2+K^2}} = \cos\varphi$, $\frac{B}{\sqrt{B^2+K^2}} = \sin\varphi$, we have, $dN/dx = \alpha cN_c\cos(\alpha x + \varphi)$. Substitute this equation into equation (21) to get,

$(2R+\zeta)\cos(\alpha x + \varphi) = \frac{3}{2c\alpha N_c}$. To evaluate $2R+\zeta$, make β = α$x$ + φ and integral the equation with β in the interval [-π/2, π/2], that is, $2R+\zeta = \frac{3\pi}{4c\alpha N_c}$.

Substitute it into the equation (18), we get,

$$q = \frac{3k\gamma\alpha^2}{\eta R^2}\left(R - \frac{\pi\cos\theta}{4c\alpha N_c}\right) \quad (22)$$

From equation (22), the critical wavenumber (at $q = 0$) is,

$$\alpha = \frac{2\pi}{\lambda} = \frac{\pi\cos\theta}{4cN_c R} \quad (23)$$

Equation (18) indicates that amplification factor $q$ increases as $\zeta$ decreases. The reduction of $\zeta$ results from the evaporation. In fact, as the air invades the space between the clusters, the drying process nearly complete so that the further evaporation time is short. Therefore, we approximately consider $\zeta$ (or $q$) as the constant, and use the critical wavenumber (at $q = 0$) to describe the fluctuation process induced by the energy criterion $W - E \leq 0$.

From equation (23), the fluctuation wavelength is,

$$\lambda = 8cN_c R/\cos\theta \quad (24)$$

When the drying has completed, the film of dried deposition consisting of $N_c$ clusters lies on the solid substrate of the area $L$. Since the substrate area $L$ is a constant for the drying experiment, cluster numbers $N_c$ increase with film thickness $H$ increasing. $R$ is constant during the process of phase separation, thus, $H \propto N_c R$. Then equation (24) results in the relation,
$$\lambda \propto H \qquad (25)$$

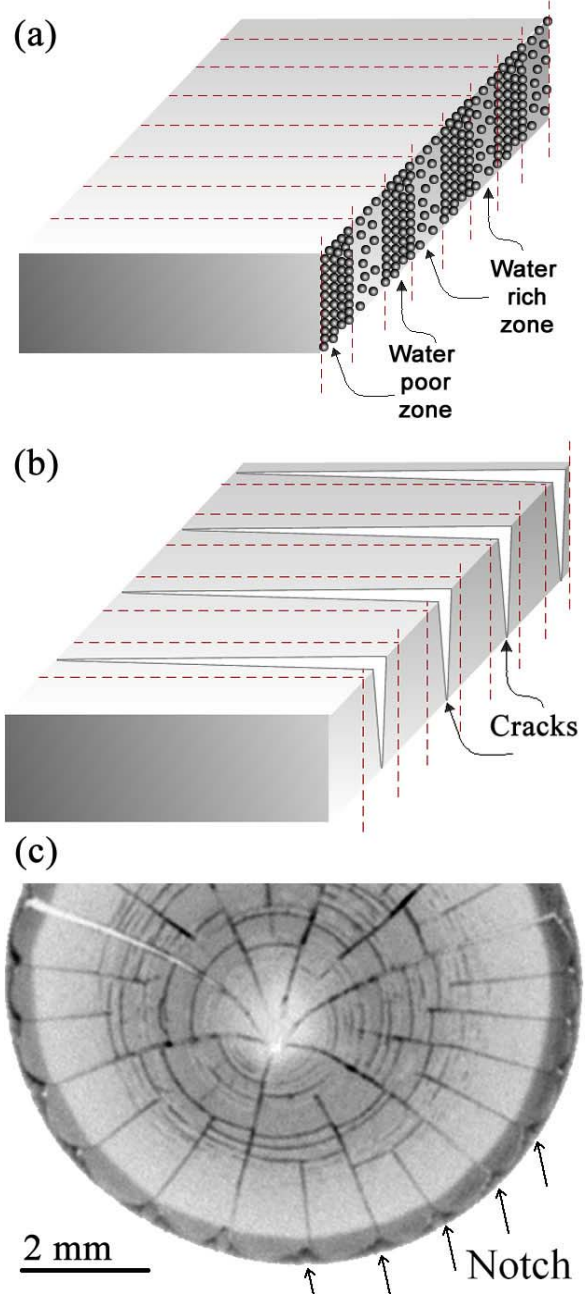

Fig. 4. (a) Fluctuation of the number concentration of clusters. (b) Cracks nucleating at the dried sites which originally are water zones due to the phase separation. (c) Crack initiation from the periodically distribution of notches resulting from the phase separation. Experimental condition see Appendix A.

As the fluctuation of cluster numbers proceeds with the time during phase separating of water and clusters, the periodical distribution of clusters forms (the cluster rich and water rich zones shown in Fig. 4(a)). The water is evaporating and finally the water zone dries completely, resulting in the periodical distribution of crevices. Flaw leading to crack pattern in thin film has been theoretically analyzed [21-23] and experimentally found [24-26]. In the present study, those crevices are the kind of periodical flaws which act as the sites where periodically parallel cracks nucleate (Fig. 4(b)). Our experimental result clearly displays that the crack pattern results from the periodical notches (flaws) formed in drying experiment (Fig. 4(c)). The cracks initiate at the crevices and then prorogate under the law of fracture mechanics. Recently, crack patterns resulting from periodical notches has also been discovered in the monolayer of hydrophobic particles at the air-water interface [27].

At a fixed time, the fluctuation of cluster numbers can be represented as a sine function (Fig. 5(a)). The fluctuation wavelength is the crack space (the distance between two adjacent cracks) in the dried film. In terms of the relation (25), the crack space linearly increases with film thickness increasing (Fig. 5(b)). This theoretical predication is consistent with the well accepted experimental results [7, 8].

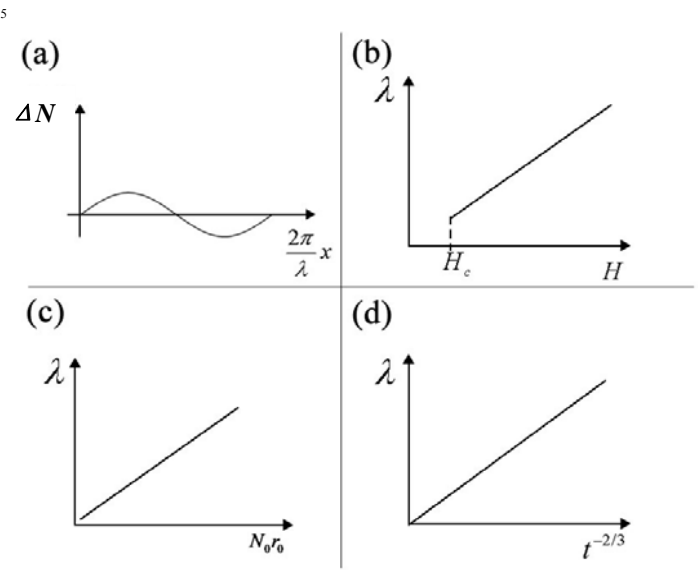

Fig. 5. Relationship between the parameters in the theory (presenting in dimensionless quantities). Fluctuation of the numbers of clusters displays the sine function (a). Crack space linearly grows with the increase of film thickness (> $H_c$, critical film thickness) (b), $N_0 r_0$ (c), $t^{-2/3}$ (d), respectively.

Combining equations (3), (4) and (24), it gives, $\lambda = 8cN_0r_0/2^{\frac{2}{3}k} \cos\theta$, thus,

$\lambda \propto N_0r_0$ or $n_0V_0r_0$  (26)

This formula indicates that the crack space increases with the increment of the numbers and size of initial colloidal particles (Fig. 5(c)). This prediction is also consistent with the experimental fact that as the particle size and suspension volume are constant, the crack spacing and thickness of deposition film linearly grow with the increase of initial particle concentration $n_0$ [4, 9].

Cracks are only produced in the deposition film whose thickness exceeds a critical value in terms of fracture mechanics [28]. There may have another reason for the production of the crack-free thin film. From the formula (25), as $H$ is very small, so does $\lambda$. Thus the fluctuation needs too high energy to occur [29], i.e. no crack (flaw) patterns produce. This may provide the explanation for the following experimental phenomenon for dried deposition film [30]: (1) Due to the lack of flaw where crack origins, the film was crack-free when its thickness was below the critical film thickness. (2) With film thickness just reaching critical film thickness, some random cracks dominated by fracture mechanics appeared, but no crack pattern produced since the corresponding $\lambda$ was still too small to develop. (3) As film thickness continuously increasing, regular crack pattern formed since suitable $\lambda$ can develop for the fluctuation.

Combining equations (3), (4), (9) and (24), results in the relation, $\lambda \propto \left(\frac{3\eta W(h_0 - h_t't)}{2k_BTPn_0h_0t}\right)^{2/3}$. Define the constant $I = \left(\frac{3\eta W}{2k_BTPn_0}\right)^{2/3}$, and consider $h_t' \ll h_0$, we have,

$\lambda \propto t^{-2/3} I \left(1 - \frac{h_t'}{h_0}t\right)^{2/3} \approx I\left(t^{-2/3} - \frac{2}{3}\frac{h_t'}{h_0}t^{1/3}\right) \approx t^{-2/3} I$.

Consequently,

$\lambda \propto t^{-2/3}$  (27)

Thus the fluctuation wavelength inversely changes with the evaporation time (Fig. 5(d)). As $t$ is large, $\lambda$ is too small to form the fluctuation. Therefore, no flaw is produced to result in the crack. This agrees with the experimental fact that crack-free thin film can be achieved for the slow drying [28].

In summary, we propose the theory for the origin of crack pattern on the deposition film produced by drying colloidal suspension. Cracks initiate from the periodically distributed flaws, making up the crack pattern. Those flaws result from the phase separation of the colloidal clusters and water. The phase separation is characterized by the fluctuation of cluster numbers with the wavelength corresponding to crack spacing. The crack spacing linearly grows with the increasing of film thickness and initial particle concentration. The theory also successfully predicts the prerequisite conditions for crack formation, which is consistent with the experimental results.

Appendix *A*

In Fig. 2(a), consider particles diffusing across the section area $s \cdot 1$ (take unit length along $x$ (Fig. 2(b)), the diffusion amount $Q_D = \int Jsdt$. The diffusion flux $J = D \cdot n/y = Dn \cdot \tan\theta_s/s = Dns_0/y_0s$, where $n$ is the particle number concentration, $y$ distance between the contact line and the reference position, $y_0$ initial distance, $s$ height of the suspension at the reference position, $s_0$ initial height, $\theta_s$ contact angle of the suspension on substrate. Diffusion constant $D$ is given by Stokes-Einstein equation, $D = k_BT/6\pi\eta R$ ($\eta$, viscosity). Consequently, $Q_D = Dns_0t_0/y_0$, where $t_0$ is the evaporation time. The solute amount in the volume $y_0s_0/2$ (take unit length along $x$ (Fig. 2(b)) is, $M = ny_0s_0/2$. $Q_D/M$ ( $= 2Dt_0/y_0^2$ ) can be estimated by our experimental data: drying the drop of silica aqueous suspension on glass substrate under ambient conditions. Particle diameter: 80 nm, weight fraction: 45%, drying time: 80s, $y_0 = 340$ μm. Get $Q_D/M = 0.7\%$. Thus $Q_D \ll M$, the diffusion from the right side of reference position to its left (the reaction area with the volume $y_0s_0/2$) can be neglected.

Appendix *B*

During the period for one coalescence, the concentration increment resulting from the evaporation is,

$\Delta C_E = \frac{n_0 S \Delta h}{S(h_0 - \Delta h)} = n_0 \frac{\Delta h}{h_0} / \left(1 - \frac{\Delta h}{h_0}\right) \approx n_0 \frac{\Delta h}{h_0}\left(1 + \frac{\Delta h}{h_0}\right)$

$\approx n_0 \Delta h / h_0$

, $\Delta h \ll h_0$, where $\Delta h$ is the height reduction resulting from the evaporation. As $\Delta h = h_t' t_k$ ($t_k$, the time for completing one coalescence), we have,

$$\Delta C_E = n_0 h'_t t_k / h_0 = n_0 t_k / t$$

where $t$ is the total evaporation time. For one coalescence, the number concentration halves in term of equation (3). The corresponding concentration reduction (represented by negative sign) is, $\Delta C_C = -C/2$, where $C$ is the suspension concentration. For the typical cluster sizes of 10 microns and the experimental data in Appendix $A$, times of coalescence is given by equation (4), $k$ = 21. The average evaporation time for one coalescence, $<t_k>$ = $t/k$ = 80 s/21 = 3.8 s. From equation (6), $\Delta C_E$ = 4.8% $n_0$ << $|\Delta C_C|$ = 50% $n_0$.

Appendix $C$

The work needed to be done for the drainage of water (from $X$ to $Y$ in Fig. 3(b)) is,

$$w_1 = \int_R^{-R} F \cdot d(-z) = -\int_R^0 4 \cdot \frac{2\pi r \gamma}{4} \cos(\varphi + \theta) dz$$
$$-\int_0^{-R} 4 \cdot \frac{2\pi r \gamma}{4} \cos(\varphi - \theta) dz$$

, where $F$ is the capillary force along $z$ direction, $\theta$ contact angle of water on the spherical cluster, $\varphi$ angle between the perpendicular and tangent lines at the contact point of the cluster with water (Fig. 3(b)), $r$ the projection radius of retreated contact line (Figs. 3(c) and 3(d)), $z = \sqrt{R^2 - r^2}$ (Fig. 3(d)). Performing the integration to get, $w_1 = \frac{8}{3} \pi \gamma R^2 \cos\theta$.

The extra work need to be done for the drainage of the water between $\zeta$ (Fig. 3(b)) is, $w_2 = \overline{F}\zeta = \frac{4}{3}\pi\gamma R\zeta \cos\theta$, where $\overline{F}$ is the average force ($w_1/2R$) during the drainage. The water drainage driven by the work $w_1 + w_2$ results in simultaneously releasing of surface energy of one cluster, $4 \times \frac{1}{4}\pi R^2 \gamma$ (water drainage of the square zone in Fig. 3(c)). Thus total work for water drainage in unit volume is,

$$W = n(w_1 + w_2) = \frac{4}{3}\pi\gamma nR(2R + \zeta)\cos\theta$$

Corresponding:
junma@nwu.edu.cn, jing@nwu.edu.cn

The authors gratefully acknowledge the support by NSFC (No. 11104218), Natural Science Basic Research Plan in Shaanxi Province of China (Program No. 2010JZ001), Scientific Research Foundation for the Returned Overseas Chinese Scholars (Shaanxi Administration of Foreign Expert Affairs, 2011).